\def\fnote#1{\footnote}

\documentstyle[12pt]{article}
\begin{document}
\medskip
\vspace{1cm}
\centerline{{\large \bf Degenerations of Sklyanin algebra
and Askey-Wilson polinomials}}
\medskip
\vspace{1cm}
\centerline{{\bf A.S.Gorsky}
}
\medskip
\centerline{Institute of Theoretical Physics,
}
\centerline{Box803,S-751 08,Uppsala,Sweden\footnote{Permanent
address:ITEP,B.Cheryomushkinskaya 25,Moscow,117259,Russia}
}
\medskip
\vspace{1cm}
\centerline{{\bf A.V.Zabrodin}
}
\medskip
\centerline{Enrico Fermi Institute, and Mathematical Disciplines Center,}
\centerline{University of Chicago, Chicago IL 60637, USA \footnote{Permanent
address:Institute of Chemical Physics,Kosygina 4,Moscow,117334,Russia}
}
\medskip
\vspace{1cm}
\centerline {UUITP-7/93 ,        HEPTH/9303026}
\vspace{2cm}
\centerline{{\bf Abstract}
}
\medskip
A new trigonometric degeneration of the Sklyanin algebra is found and the
functional realization of its representations in space of polynomials in
one variable is studied. A further contraction gives the standard quantum
algebra $U_{q}(sl(2))$. It is shown that the degenerate Sklyanin algebra
contains a
subalgebra isomorphic to algebra of functions on the quantum sphere
$(SU(2)/SO(2))_{q^{1\over2}}$. The diagonalization of general quadratic form in
its generators leads in the functional realization to the difference equation
for Askey-Wilson polynomials.
\par
\medskip
\newpage
\par
\medskip
Sklyanin algebra (SA) [1] plays a very important role in the theory of
quantum integrable systems being the algebra of observables in the models
with elliptic R-matrix. It is the algebra with quadratic relations that follow
from the Yang-Baxter equation for quantum monodromy matrices. The trigonometric
degeneration of the R-matrix gives rise to a number of the well-known exactly
solvable models like XXZ spin chain, sine-Gordon, etc. In this case the SA
degenerates [2] to the quantum group $ U_{q}(sl(2))$ - deformation of the

universal enveloping algebra of $ sl(2)$ introduced by Kulish and Reshetikhin
[3]
and then extensively studied in more general mathematical context [4].

\par
In this Letter, we show that $U_{q}(sl(2))$ is a very particular case of a
more general family of trigonometric degenerations of the SA. More precisely,
the standard $U_{q}(sl(2)) $can be obtained as further contraction of the
degenerate SA. On the other hand, we show that the trigonometric limit of SA
contains a subalgebra isomorphic to $ Fun_{q^{1\over2}}(SU(2)/SO(2))$ - the
algebra of
functions on the quantum sphere$ (SU(2)/SO(2))_{q^{1\over2}}.$ The generator
sof this
algebra can be realized as difference operators acting on polynomials
in one variable. This realization follows directly from the representation
theory of SA [1]. We argue that the diagonalization problem for general
quadratic form in the generators (which is commonly interpreted as the
Hamiltonian of a proper physical system) is equivalent to the difference
Hamiltonian of a proper physical system) is equivalent to the differtence
equation for the Askey-Wilson polynomials [5]. In particular, the
diagonalization of the "$SO(2)$-invariant" element of
$Fun_{q^{1\over2}}(SU(2)/SO(2))$
(i.e. the generator of the commutative algebra

$Fun_{q^{1\over2}}({SO(2)}{\setminus}{SU(2)}{/}{SO(2)})$ leads to the
Rogers-Askey-Ismail

polynomials [6]. At the same time these polynomials are known to give zonal
spherical functions on the quantum spheres [7,8].

\par
The SA is the two-parametric deformation of $ U_{q}(gl(2))$. It is the
algebra with 4 generators $ S_{0}, S_{1}, S_{2}, S_{3}$ and the following
quadratic relations [1]:
\par
\begin{eqnarray}
 [S_{0},S_{\lambda}]=iJ_{\mu\nu}(S_{\mu},S_{\nu})\nonumber\\
 {[S_{\lambda},S_{\mu}]}=i(S_{0},S_{\nu})

\end{eqnarray}
 Here [,] and (,) are commutator and anticommutator respectively and

$(\lambda, \mu, \nu)$ stands for any cyclic permutation of $ (1,2,3)$. The
structure
constants $J_{\mu \nu}$ are parametrized as follows: $J_{\mu \nu}=(J_{\mu}-
J_{\lambda})/J_{\nu}$ where
\par
\begin {eqnarray}
 J_{1}=\theta_{4}(2\gamma)\theta_{4}(0)/(\theta_{4}(\gamma))^{2}\nonumber\\
 J_{2}=\theta_{3}(2\gamma)\theta_{3}(0)/(\theta_{3}(\gamma))^{2}\\

 J_{3}=\theta_{2}(2\gamma)\theta_{2}(0)/(\theta_{2}(\gamma))^{2}\nonumber

\end{eqnarray}
\par
 We use the standard notation [9] for the theta functions with elliptic nome
$h=exp(i\pi\tau)$. The two deformation parameters are just h and $\gamma$ they
are supposed to be real. The two central (Casimir) elements are given by
\par
\begin{eqnarray}
K_{0}=(S_{0})^{2}+(S_{1})^{2}+(S_{2})^{2}+(S_{3})^{2}\\

K_{1}=J_{1}(S_{1})^{2}+J_{2}(S_{2})^{2}+J_{3}(S_{3})^{2}\nonumber

\end{eqnarray}
 \noindent  Trigonometric degeneration of the SA can be obtained by tending h
to 0.
To do this let us redefine the generators as follows:
\par
\begin{eqnarray}
A=ih^{-1\over4}(S_{0}-tanh(\pi\gamma)S_{3})/(4sinh(\pi\gamma))\nonumber\\

D=ih^{-1\over4}(S_{0}+tanh(\pi\gamma)S_{3})/(4sinh(\pi\gamma))\\

C=ih^{+1\over4}(S_{1}-iS_{2})/(2sinh(2\pi\gamma))\nonumber\\

B=ih^{-3\over4}(S_{1}+iS_{2})/(8sinh(2\pi\gamma))\nonumber

\end{eqnarray}
\noindent We consider only the representations of series a) in terminology
[1].It follows
from the realization of the generators of SA by difference
operators [1] that these combinations have a finite limit when h goes to 0.
Substituting (4) into (1) and denoting $ q=exp(-2\pi\gamma) $ one obtains the
relations in the degenerate SA:
\par
\begin{eqnarray}
DC=qCD,  CA=qAC\\

{[A,D]}={1\over4}(q-q^{-1})^{3}C^{2}\\

{[B,C]}=(A^{2}-D^{2})/(q-q^{-1})\\

AB-qBA=qDB-BD=-({1\over4})(q^{2}-q^{-2})(DC-CA)

\end{eqnarray}
\noindent We would like to stress that though $ J_{12}=0 $ when $ h=0 $
and naively A and D would commute the careful limiting procedure taking into
account the
order
in h of each term in (1) gives just the nonvanishing commutator (6).
\par
According to the general representation theory of the SA developed in [1]
the finite dimensional irreducible representations of the limiting algebra
(5)-(8) are parametrized by a non-negative integer or half-integer number j
(spin of the representation). The spin-j representation can be realized in

(2j+1)-dimensional space of polynomials of degree 2j by the following
difference
operators:
\par
\begin{eqnarray}
C=({2\over(q-q^{-1})})({1\over(z-z^{-1})})(T_{+}-T_{-})\\

A=q^{-j}({1\over(z-z^{-1})})(zT_{+}-z^{-1}T_{-})\\

D=q^{j}({1\over(z-z^{-1})})(zT_{-}-z^{-1}T_{+})\\

B=({1\over(2(q-q^{-1})})({1\over(z-z^{-1})})(q^{2j}(z^{2}T_{-}-z^{-2}T_{+})-\\
  -q^{-2j}(z^{2}T_{+}-z^{-2}T_{-}) - (q+q^{-1})(T_{+}-T_{-})) \nonumber

\end{eqnarray}
\noindent where $T_{+}f(z)=f(qz), T_{-}f(z)=f(q^{-1}z)$. In the classical limit
$q=1$
(9-12) give the standard functional realization of the $sl(2)$-generators
C,B and $ H=(D-A)/(4\pi\gamma)$ by differential operators.
\par

It is remarkable that the standard quantum algebra $U_{q}(sl(2))$ can be
obtained from (5-8) by the contraction procedure: B to B, A to ${\epsilon}A$,
D to ${\epsilon}D$, C to $ {\epsilon}^{2}C$;$\epsilon \rightarrow 0$.
 For the representation (9-12) the contraction effectively means

that z goes to infinity and we can formally put $z^{-1}$ equal to 0
thus obtaining the well-known realization of $U_{q}(sl(2)) $in space of
holomorphic functions.
\par

One can see from (3) that the trigonometric limit of the central element
$K_{0}-K_{1}$ is proportional to $ AD+(1/4q)(q-q^{-1})^{2}C^{2}$. It follows
from (9-12) that the value of this Casimir element does not depend on j and
is equal to 1. Therefore, this operator can be put equal to 1:
\par
\begin{equation}
 AD+{1\over(4q)}(q-q^{-1})^{2}C^{2}=1

\end{equation}
\noindent This relation is to be added to (5-8).
\par

   Note that A,D,C generate the subalgebra with the relations (5),(6),(13).
Remarkably enough, it is isomorphic to the algebra of functions on the

quantum sphere $(SU(2)/SO(2))_{q^{1\over2}}$ [8] ( note this change of q to

$q^{1\over2}$!). Let us point out that the generator A+D $(the former S_{0})$
 corresponds under this isomorphysm to the simplest nontrivial element of

 $Fun_{q^{1\over2}}(SU(2)/SO(2))=F_{q^{1\over2}}$ invariant with respect to the
quasiregular action of the proper "twisted primitive element" [7] of
$U_{q^{1\over2}}(su(2))$ representing the "infinitesimal $SO(2)$-rotation".

In other words, the (commutative) algebra of functions on the double quantum

coset
space $ Fun_{q^{1\over2}}({SO(2)}{\setminus}{SU(2)}/{SO(2)})$ is generated by 1
and
A+D. The

operators (9-11) form a (reducible) representation of $F_{q^{1\over2}}$
\par

Now let us turn to the problem of diagonalization of a quadratic form
in generators of $F_{q^{1\over2}}$ in the space of spin-j representation of the
whole algebra (5-8). Such quadratic operator could be interpreted as the
Hamiltonian of a proper physical system. It is known [10,11] that in the
classical $sl(2)$ case all exactly solvable problems of quantum mechanics on
the line can be obtained in this way.
\par

   Consider the general quadratic form in A,D,C:
\par
\begin{eqnarray}
Q(r,k;\alpha,\beta)=q^{r}A^{2}+q^{-r}D^{2}-{1\over4}(q^{2k+1}+ \\
+q^{-2k-1})(q-q^{-1})^{2}C^{2}+(q-q^{-1})({\alpha}AC+{\beta}DC)\nonumber
\end{eqnarray}

\noindent where $r,k,\alpha,\beta$ are arbitrary parameters. Substituting
(9-11) into (14)
we obtain the following eigenvalue equation for Laurent polynomials in z
$P_{m}(z)$:
\par
\begin{equation}
Q(r,k;\alpha,\beta)P_{m}(z)=E_{m}P_{m}(z)

\end{equation}
\noindent One can see that when $E_{m}=q^{2(j-m)-r}+q^{-2(j-m)+r} $ (15)
coincides with
the difference equation for Askey-Wilson polynomials

$P_{m}({(z+z^{-1})\over2}; a,b,c,d|q^{2})$ (=$P_{m}(z)$ for brievity)
\par
\begin{eqnarray}
A(z)(P_{m}(q^{2}z)-P_{m}(z)) + A(z^{-1})(P_{m}(q^{-2}z)-P_{m}(z))=\\

= (q^{-2m}-1)(1-abcdq^{2m-1})P_{m}(z)\nonumber

\end{eqnarray}
\noindent where
$A(z)=(1-az)(1-bz)(1-cz)(1-dz)/((1-z^{2})(1-q^{2}z^{2}))$
and the parameters a,b,c,d are expressed through $ r,k,\alpha,\beta $ as
follows:
\par
\begin{eqnarray}
(1-az)(1-bz)(1-cz)(1-dz)=1+{2\beta}q^{r+1-j}-\\

-(q^{2k+1}+q^{-2k-1})q^{r+1-2j}z^{2}+{2\alpha}q^{r+1-3j}z^{3}+q^{2r+2-4j}z^{4}\nonumber
\end{eqnarray}
\noindent  In particular, for $ r=k=\alpha=\beta=0 $ we have $
Q(0,0,0,0)=(A+D)^{2}-2$
and (16)
reduces to the equation for the Rogers-Askey-Ismail polynomials
$C_{m}((z+z^{-1})/2; q^{-2j}|q^{2})$ [6] which are also known as Macdonald's
polynomials for the root system $A_{1} $[12]. So the eigenfunctions of the
$SO(2)$-invariant element of $F_{q^{1\over2}}$ discussed above are expressed in
terms
of these polynomials. We note also that at the same time the Askey-Wilson
polynomials (with particular values of the parameters) provide the most
general family of zonal spherical functions on the quantum group
$U_{q}(su(2))$ [7].
\par
In conclusion let us make a few remarks. The diagonalization of a more
general quadratic form in all the generators including B leads to a more
complicated eigenvalue difference equation corresponding to the so-called
quasi-exactly solvable problems [11] in the classical limit. In case of the
contraction to $ U_{q}(sl(2))$ the little Jacobi polynomials appear.
\par
The role of the  quadratic algebra (5-8) (obtained as a result of the
limiting procedure from the SA) in integrable models with trigonometric
R-matrix is not clear enough at the moment. However, as it was shown in
[13] the scatetring of dressed excitations in the XXZ antiferromagnetic spin
chain can be described in terms of zonal spherical elements of the algebra
of functions on the quantum sphere and/or hyperboloid.

\par
   We wish to thank P.G.O.Freund and M.A.Olshanetsky for illuminating
discussions. A.Z. is grateful to the Mathematical Disciplines Center,
University of Chicago where this work was completed for the hospitality
and support.A.G. thanks  A.Niemi for the warm hospitality in the Institute of
Theoretical
Physics .

\par

\par
\end{document}